\documentclass[floatfix,aps,nofootinbib,superscriptaddress,twocolumn]{revtex4}

\pdfoutput=1
\usepackage{graphicx}
\usepackage{epstopdf}
\usepackage{amsmath}
\usepackage{amsfonts}
\usepackage{amssymb}
\usepackage{color}
\usepackage{mathrsfs}
\usepackage{multirow}
\usepackage{bm}
\usepackage[colorlinks=true, linkcolor=red, citecolor=blue]{hyperref}

\def\({\left(}
\def\){\right)}
\def\[{\left[}
\def\]{\right]}

\def\e{\begin{equation}}
\def\q{\end{equation}}
\def\m{\begin{eqnarray}}
\def\n{\end{eqnarray}}
\def\ttp{\texttt{Planck }}

\usepackage{acro}
\DeclareAcronym{CL}{
	short = CL ,
	long  = confidence level
}

\DeclareAcronym{CMB}{
	short = CMB ,
	long  = cosmic microwave background
}

\DeclareAcronym{BAO}{
	short = BAO ,
	long  = baryon acoustic oscillation
}

\DeclareAcronym{LSS}{
	short = LSS ,
	long  = large scale structure
}

\DeclareAcronym{eBOSS}{
	short = eBOSS,
	long = extended Baryon Oscillation Spectroscopic Survey
	}

\usepackage[capitalise]{cleveref}
\crefname{figure}{Fig.}{Figs.}
\Crefname{figure}{Fig.}{Figs.}

\begin{document}

\title{Constraints on the sum of neutrino masses using cosmological data including the latest extended Baryon Oscillation Spectroscopic Survey DR14 quasar sample}

\author{Sai Wang}
\email{wangsai@itp.ac.cn}
\affiliation{Department of Physics, The Chinese University of Hong Kong, Shatin, N.T., Hong Kong SAR 999077, China}
\author{Yi-Fan Wang}
\email{yfwang@phy.cuhk.edu.hk}
\affiliation{Department of Physics, The Chinese University of Hong Kong, Shatin, N.T., Hong Kong SAR 999077, China}
\author{Dong-Mei Xia}
\email{xiadm@cqu.edu.cn}
\affiliation{Key Laboratory of Low-grade Energy Utilization Technologies \& Systems of Ministry of Education of China, College of Power Engineering, Chongqing University, Chongqing 400044, China}

\begin{abstract}
We investigate the constraints on the sum of neutrino masses ($\Sigma m_\nu$) using the most recent cosmological data, which combines the distance measurement from baryonic acoustic oscillation in the extended Baryon Oscillation Spectroscopic Survey DR14 quasar sample with the power spectra of temperature and polarization anisotropies in the cosmic microwave background from the Planck 2015 data release. We also use other low-redshift observations including the baryonic acoustic oscillation at relatively low redshifts, the supernovae of type Ia and the local measurement of Hubble constant. In the standard cosmological constant $\Lambda$ cold dark matter plus massive neutrino model, we obtain the $95\%$ \acl{CL} upper limit to be $\Sigma m_\nu<0.129~\mathrm{eV}$ for the degenerate mass hierarchy, $\Sigma m_{\nu}<0.159~\mathrm{eV}$ for the normal mass hierarchy, and $\Sigma m_{\nu}<0.189~\mathrm{eV}$ for the inverted mass hierarchy. Based on Bayesian evidence, we find that the degenerate hierarchy is positively supported, and the current data combination can not distinguish normal and inverted hierarchies. Assuming the degenerate mass hierarchy, we extend our study to non-standard cosmological models including the generic dark energy, the spatial curvature, and the extra relativistic degrees of freedom, respectively, but find these models not favored by the data.


\end{abstract}

\maketitle

\acresetall

\section{Introduction}
The phenomena of neutrino oscillation have provided convincing evidence for the neutrino non-zero masses and the mass splittings (see Ref.~\cite{Olive:2016xmw} for a review). 
However,  the current experimental results can not decisively tell if the third neutrino is heavier than the other two or not. 
There are thus two potential mass hierarchies for three active neutrinos, namely, the normal hierarchy (NH), in which the third neutrino is the heaviest, and the inverted hierarchy (IH), in which the third neutrino is the lightest. 
The sum of neutrino masses ($\Sigma m_{\nu}$) is also remained unknown, and different neutrino hierarchies would have different total mass. 
Taking into account the experimental results of the square mass differences \cite{Olive:2016xmw}, the lower bound on $\Sigma m_{\nu}$ is estimated to be $0.06\textrm{eV}$ for NH, while to be $0.10\textrm{eV}$ for IH. 
The upper limit on $\Sigma m_{\nu}$ is much loose based on the experimental particle physics. 
For instance, the most sensitive neutrino mass measurement to date, involving the kinematics of tritium beta decay, provides a $95\%$ \ac{CL} upper bound of $2.05\textrm{eV}$ on the electron anti-neutrino mass \cite{Aseev:2011dq}.

Cosmology plays a significant role in exploring the neutrino masses (see Ref.~\cite{Lesgourgues:2006nd} for a review), since they can place stringent upper limits on $\Sigma m_{\nu}$, which is a key to resolving the neutrino masses by combining with the square mass differences measured. 
Massive neutrinos are initially relativistic in the Universe, and become non-relativistic after a transition when their rest masses begin to dominate. 
Imprints have been left on the cosmic microwave background (CMB) and the \ac{LSS}. 
The CMB is affected through the early-time integrated Sachs-Wolfe effect \cite{Hou:2012xq}, which shifts the amplitude and location of the CMB acoustic peaks due to a change of the redshift of matter-radiation equality. 
The LSS is modified through suppressing the clustering of matter, due to the large free-streaming velocity of neutrinos. Therefore, the massive neutrinos can be weighed using the measurements of CMB and LSS \cite{Hu:1997mj,Hu:2001bc,Komatsu:2008hk,Ade:2015xua}.

Assuming the $\Lambda$ cold dark matter ($\Lambda$CDM) model and three active neutrinos are in degenerate hierarchy (DH) , namely with equal mass, the \ttp  Collaboration recently reported the $95\%$ \ac{CL} upper limit on $\Sigma m_\nu$ to be $0.49\textrm{eV}$ using the CMB temperature and polarization anisotropies from the \ttp 2015 data \cite{Ade:2015xua}.
Adding the gravitational lensing of the CMB from \ttp 2015 data relaxes the upper limit to be $0.59\textrm{eV}$, a less stringent one. 
Given the accuracy of the \ttp 2015 data, assuming the neutrinos in NH and IH would have negligible impact on the constraints on the sum of neutrino masses.

Combining \ac{BAO} and other low redshift data such as local Hubble constant $H_0$ and Type Ia supernovae can further tighten the constraints \cite{Reid:2009nq,Thomas:2009ae,Riemer-Sorensen:2013jsa,Swanson:2010sk,Cuesta:2015iho,Palanque-Delabrouille:2015pga,DiValentino:2015sam,Rossi:2014nea,Xu:2016ddc,Jimenez:2010ev}. There are also efforts on the constraints by combining experimental data from particle physics \cite{Capozzi:2017ipn,Caldwell:2017mqu,Hannestad:2016fog}.
Especially, since the \ac{BAO} data can significantly break the acoustic scale degeneracy, the sum of neutrino masses is tightly constrained to $0.15\textrm{eV}$ by Ref.~\cite{Huang:2015wrx} after adding the \ac{BAO} data from \ac{LSS} surveys \cite{Beutler:2011hx,Ross:2014qpa,Gil-Marin:2015nqa}. 
This result is  close to the lower bound of neutrino masses in the IH, namely 0.10eV.
Current tightest constraints on the neutrino total mass \cite{Vagnozzi:2017ovm,Giusarma:2016phn} from cosmological data, especially from \ttp high-$l$ polarization data of CMB, already reached $\sim 0.10\textrm{eV}$ given certain combinations of data set, implying a favor of NH for neutrino. Nevertheless, it is necessary to take the mass prior set by the neutrino mass hierarchy into account to get consistent constraints. 
After taking the square mass differences into account, we \cite{Huang:2015wrx,Wang:2016tsz} found that the upper limit of neutrino total mass becomes $0.18\textrm{eV}$ for NH, and $0.20\textrm{eV}$ for IH. 
It was also shown by Ref.~\cite{Wang:2016tsz,Zhang:2015uhk,Allison:2015qca,DiValentino:2015wba,Gerbino:2015ixa,Zhao:2016ecj,Gerbino:2016sgw,Li:2017iur,Yang:2017amu} that an extension of the standard  cosmology model, \textit{e.g.}, dynamical dark energy and non-zero spatial curvature, would have subtle impacts on constraining the sum of neutrino masses.
In this work we investigate the mass prior effect on constraining the neutrino total mass in the standard cosmological model.

Most recently, the SDSS-IV \ac{eBOSS} \cite{Ata:2017dya} measured a \ac{BAO} scale in redshift space in the redshift interval $1<z<2$  for the first time, using the clustering of 147,000 quasars with redshift $0.8<z<2.2$. A spherically averaged \ac{BAO} distance  to $z=1.52$ is obtained, namely,
$D_{V}(z=1.52)=3843\pm147(r_d/r_{d,\textrm{fid}})~\textrm{[Mpc]}$, 
which is of $3.8\%$ precision. 
In this paper, this data point is denoted by \texttt{eBOSS DR14} for simplicity. 
The SDSS-IV \ac{eBOSS} measurement of the \ac{BAO} scale is expected to break the degeneracy between the NH and IH scenarios of three active neutrinos at $2\sigma$ confidence level \cite{Zhao:2015gua}. 

In this work, we constrain the sum of neutrino masses in the $\Lambda$CDM model by adding the recently released \texttt{eBOSS DR14} data.
Besides the maximal likelihood analysis, we also employ the Bayesian statistics  to infer the parameters, and especially to perform model selections. 
We expect to show that the current observations can improve the previous constraints on $\Sigma m_\nu$ significantly, and the neutrino mass hierarchies have to be considered to analyze the current observational data.	
Due to possible degeneracy between the sum of neutrino masses and a few extended cosmological parameters, similar studies are proceeded under the framework of extended cosmological models by introducing the generic dark energy ($w$), the non-zero spatial curvature ($\Omega_k$), and the extra relativistic degree of freedom ($N_{\textrm{eff}}$), respectively.

The rest of this paper is arranged as follows. Sec.~\ref{sec:method} introduces the adopted cosmological models, the cosmological observations, and the method of statistical analysis. Sec.~\ref{sec:result1} shows the result of the constraints on $\Sigma m_\nu$ in the $\Lambda$CDM model, while Sec.~\ref{sec:result2} shows the effect of a few extended cosmological parameters on the constraints on $\Sigma m_\nu$. In Sec.~\ref{sec:summary}, the conclusions are summarized.

\section{Models, dataset \& methodology}\label{sec:method}
\subsection{Cosmological models}\label{subsec:model}
In the $\Lambda$CDM plus massive neutrino model (hereafter, $\nu\Lambda$CDM for short), we put constraints on the sum of neutrino masses with and without taking into account the neutrino mass hierarchies. 
Specifically, we consider the DH, NH, and IH of the neutrinos. Based on the neutrino oscillation phenomenon, the square mass differences between three active neutrinos have been measured to be $\Delta m_{21}^2=7.5\times10^{-5}\textrm{eV}^2$ and $|\Delta m_{31}^2|=2.5\times10^{-3}\textrm{eV}^2$ \cite{Olive:2016xmw}. Therefore, there is a lower bound, i.e. $\Sigma m_\nu\geq0.06\textrm{eV}$, for the NH, and $\Sigma m_\nu\geq0.10\textrm{eV}$ for the IH. 
There is not such a lower bound for the DH, but $\Sigma m_\nu$ should deserve a positive value. 

In the  $\nu\Lambda$CDM model, there are six base parameters denoted by $\{\omega_b, \omega_c, 100\theta_{\textrm{MC}}, \tau, n_s, \textrm{ln}(10^{10}A_s)\}$ plus a seventh independent parameter denoted by $\Sigma m_{\nu}$ for the sum of neutrino masses. 
Here $\omega_b$ and $\omega_c$ are, respectively, physical densities of baryons and cold dark matter today. $\theta_{\textrm{MC}}$ is the ratio between sound horizon and angular diameter distance at the decoupling epoch. $\tau$ is Thomson scatter optical depth due to reionization. $n_s$ and $A_s$ are, respectively, spectral index and amplitude of the power spectrum of primordial curvature perturbations. The pivot scale is set to be $k_{p}=0.05\textrm{Mpc}^{-1}$.

When the generic dark energy is taken into account, an eighth independent parameter is introduced, which describes the equation of state (EoS) of the dark energy. This parameter is denoted by $w$, and the corresponding cosmological model is the $\nu w$CDM model. When the spatial curvature is considered, the eighth independent parameter is denoted by $\Omega_k$, and the corresponding model is the $\nu\Omega_k\Lambda$CDM model. When the extra relativistic degree of freedom is considered, the eighth independent parameter is denoted by $N_{\textrm{eff}}$, and the corresponding model is the $\nu N_{\textrm{eff}} \Lambda$CDM model. 
For the above three extended models, we only consider the upper limits on $\Sigma m_\nu$ in the DH scenario of three massive neutrinos, because the neutrino mass hierarchy would have negligible effects on the constraints given the current cosmological observations.

\subsection{Cosmological data}
The cosmological observations adopted by this work include CMB, BAO, and other low-redshift surveys. To be specific, the CMB data are composed of temperature anisotropies, polarizations, and gravitational lensing of the CMB reported by the \ttp 2015 data release \cite{Ade:2015xua}. 
The CMB lensing is used here since it is very sensitive to the neutrino masses \cite{Lewis:2006fu}.
Specifically, we utilize the angular power spectra of TT, TE, EE, lowTEB, and gravitational lensing of the CMB. The BAO data points come from the \texttt{6dF galaxy survey} \cite{Beutler:2011hx}, \texttt{SDSS DR7} main galaxy sample \cite{Ross:2014qpa}, \texttt{SDSS-III BOSS DR12 LOWZ} and \texttt{CMASS} galaxy samples \cite{Gil-Marin:2015nqa}, and the \texttt{SDSS-IV eBOSS DR14} quasar sample \cite{Ata:2017dya}. 
The supernovae dataset is the ``joint light-curve analysis'' (JLA) compilation of the supernovae of type Ia (SNe Ia) \cite{Betoule:2014frx}. The local measurement of Hubble constant ($H_0$) comes from the \texttt{Hubble Space Telescope} \cite{Riess:2016jrr}. The full data combination combines together all the cosmological observational data mentioned above.
In fact, one can further add other astrophysical data, such as the galaxy weak lensing \cite{Heymans:2012gg,Erben:2012zw}, the redshift space distortion \cite{Samushia:2013yga}, and the \texttt{Planck} cluster counts \cite{Ade:2013lmv}, to improve the constraints on the neutrino masses.
Though these datasets are directly related to the neutrino masses, however, the amplitudes of power spectra of the cosmological perturbations obtained from these observations are in tension with the one obtained from the \texttt{Planck} CMB data. It is believed that these observations deserve underlying uncontrolled systematics, which may bias the global fitting. Hence, we do not take them into account in this paper.

\subsection{Statistical method}
Given the observational dataset and the corresponding likelihood functions, we utilize the Markov-Chain Monte-Carlo (MCMC) sampler in the CosmoMC \cite{Lewis:2002ah} to estimate across the parameter space, and the PolyChord \cite{Handley:2015fda,Handley:201506} plug-in of the CosmoMC to calculate the Bayesian evidence for model selection.

The Bayesian evidence ($E$) is defined as an integral of posterior probability distribution function (PDF), i.e. $P(\theta)$, over the parameter space \{$\theta$\}, i.e. $
E=\int d\theta P(\theta)$  \cite{BayesFactor:1995}.
Given two different models $M_1$ and $M_2$, the logarithmic Bayesian factor is evaluated as $
\Delta \ln E=\ln E_{M_1}-\ln E_{M_2}$.
When $0<\Delta \ln E<1$, the given dataset indicates no significant support for either model. When $1<\Delta \ln E<3$, there is a positive support for $M_1$. When $3<\Delta \ln E<5$, there is a strong support for $M_1$. When $\Delta \ln E>5$, there is a very strong support for $M_1$. 
Conversely, the negative values mean that the dataset supports $M_2$, rather than $M_1$. 

In this work, $M_2$ usually denotes $\nu_{\mathrm{DH}}\Lambda$CDM while $M_1$ denotes one of other models. An exception is that $M_1$ denotes the NH while $M_2$ denotes the IH, when we compare the NH and the IH in $\nu\Lambda$CDM. The parameter space $\{\theta\}$ is consist of the six base parameters plus the sum of neutrino masses for $\nu\Lambda$CDM, while an eighth parameter is further added when an extended model is considered. For each model, the independent parameters have been showed explicitly in section~\ref{subsec:model}.

We also evaluate the best-fit $\chi^2$. For any scenario, a smaller value of the best-fit $\chi^2$ implies that this scenario fits the dataset better. For both Bayesian and maximal likelihood methods, the prior ranges for all the independent parameters are set to be sufficiently wide to avoid affecting the results of data analysis.

\section{Results for the $\nu\Lambda$CDM model}\label{sec:result1}

\begin{figure}[htbp]
\includegraphics[width=1\columnwidth,height=1\columnwidth]{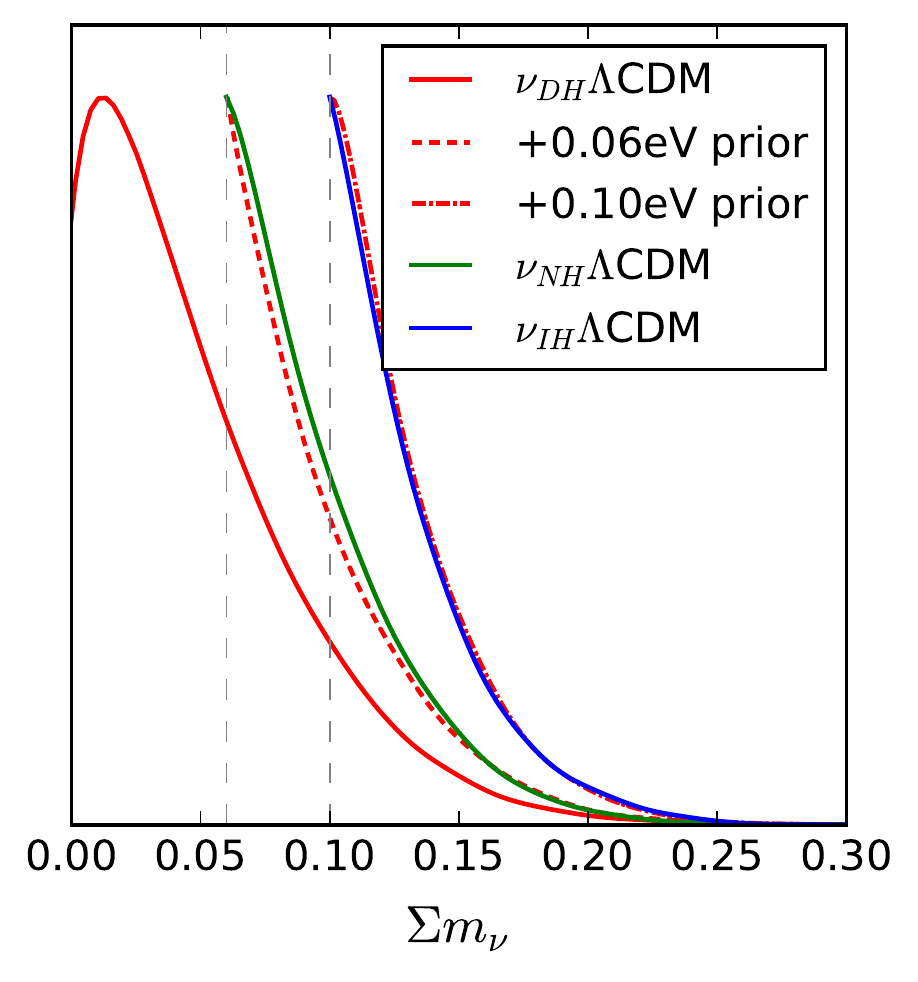}
\caption{The posterior probability distribution functions of the sum of neutrino masses for three mass hierarchies of massive neutrinos. The red, green, and blue solid curves denote DH, NH, and IH, respectively. The red dashed (dot-dashed) curve denotes DH with a lower bound of $0.06\textrm{eV}$ ($0.10\textrm{eV}$) on $\Sigma m_\nu$. From left to right, the vertical dashed lines denote $\Sigma m_\nu=0.06~\mathrm{eV},~0.10~\mathrm{eV}$, respectively.}
\label{fig:mnu}
\end{figure}

For the $\nu\Lambda$CDM model, we present the results of parameter inference and model comparison in Tab.~\ref{tab:para} and in Fig.~\ref{fig:mnu}. To be specific, we show the $68\%$ \ac{CL} constraints on the six base parameters of the $\Lambda$CDM, and the $95\%$ CL upper limits on the sum of neutrino masses in Tab.~\ref{tab:para}. For three mass hierarchies of massive neutrinos, we find that the constraints on the six base parameters of the $\Lambda$CDM are compatible within $68\%$ CL. We also list in Tab.~\ref{tab:para} the best-fit values of $\chi^2$ and the logarithmic Bayesian evidences. In Fig.~\ref{fig:mnu}, we depict the posterior PDFs of the sum of neutrino masses. The red, green, and blue solid curves, respectively, denote the posterior PDFs of $\Sigma m_\nu$ for DH, NH, and IH of massive neutrinos. In addition, we wonder if the difference in neutrino mass constraints are due to the different priors in the parameter $\Sigma m_\nu$. Therefore, we also depict in Fig.~\ref{fig:mnu} the neutrino mass constraints for the $\nu_{\textrm{DH}}\Lambda$CDM model with two non-vanishing lower bounds, i.e., $0.06\textrm{eV}$ (red dashed curve) and $0.10\textrm{eV}$ (red dot-dashed curve).

\begin{table}[htbp]
\centering
\renewcommand{\arraystretch}{1.5}
\scalebox{0.89}[0.89]{%
\begin{tabular}{|c|c|c|c|c|}
\hline
& $\nu_{\textrm{DH}}\Lambda$CDM & $\nu_{\textrm{NH}}\Lambda$CDM & $\nu_{\textrm{IH}}\Lambda$CDM  \\
\hline
$\omega_b$ & 
 $0.02238\pm0.00014$ & $0.02240\pm0.00014$ & $0.02242\pm0.00014$ \\
$\omega_c$ & $0.1178\pm0.0010$ & $0.1174\pm0.0010$ & $0.1171\pm0.0010$ \\
$100\theta_{\textrm{MC}}$ &  $1.04105\pm0.00030$ & $1.04107\pm0.00029$ & $1.04108\pm0.00030$ \\
$\tau$ & $0.0708\pm0.0133$ & $0.0775\pm0.0131$ & $0.0825\pm0.0128$ \\
$n_s$ & $0.9692\pm0.0040$ & $0.9704\pm0.0041$ & $0.9711\pm0.0040$ \\
$\textrm{ln}(10^{10}A_s)$ & $3.071\pm0.025$ & $3.084\pm0.024$ & $3.093\pm0.024$ \\
$\Sigma m_\nu~[\textrm{eV}]$ & $<0.129$ & $<0.159$ & $<0.189$ \\
\hline
$\chi^2_{\textrm{min}}/2$ & $6832.461$ & $6833.353$ & $6833.577$ \\
$\ln E$ & $-6890.50\pm0.23$ & $-6892.61\pm0.23$ & $-6892.54\pm0.23$ \\
\hline
\end{tabular}}
\caption{The $68\%$ CL constraints on six base parameters of the $\Lambda$CDM, the $95\%$ CL upper limits on the sum of neutrino masses, as well as the best-fit values of $\chi^2$ and the logarithmic Bayesian evidence, i.e. $\ln E$ ($68\%$ CL).}
\label{tab:para}
\end{table}

For three neutrinos with degenerate mass, the upper limit on the sum of neutrino masses is obtained to be
\begin{equation}\label{eq:dh}
\Sigma m_\nu<0.129~\textrm{eV}~~~ (95\%~ \textrm{CL}).
\end{equation}
This upper limit is close to the lower bound of $0.10\textrm{eV}$ required by the IH scenario. In fact, the upper limit even becomes $\Sigma m_\nu<0.10~\mathrm{eV}$ if the gravitational lensing of the CMB is discarded in the global fitting. 
As expected, the upper limit $\Sigma m_\nu<0.129~\textrm{eV}$
is $3.7\%$ tighter than the existing one, e.g. $\Sigma m_\nu<0.134~\textrm{eV}$ in Wang et al.~\cite{Wang:2016tsz}, which did not include the eBOSS DR14 data, and $\Sigma m_\nu<0.197~\textrm{eV}$ in Zhang~\cite{Zhang:2015uhk}, which used a different BAO dataset.
However, this constraint is slightly looser than that of $\Sigma m_\nu<0.12\textrm{eV}$ obtained by combining BOSS Lyman-$\alpha$ with Planck CMB \cite{Palanque-Delabrouille:2015pga}.


The above results reveal the necessity of taking into account the square mass differences between three massive neutrinos when one constrains the neutrino masses with current cosmological observations. 
For three neutrinos with NH, the upper limit on the sum of neutrino masses is given by
\begin{equation}
\label{eq:nh}
\Sigma m_\nu<0.159~\textrm{eV}~~~(95\%~ \textrm{CL}),
\end{equation}
while for IH, the result is given by
\begin{equation}
\label{eq:ih}
\Sigma m_\nu<0.189~\textrm{eV}~~~(95\%~ \textrm{CL}).
\end{equation}
Here we further consider the Bayesian model selection.
The logarithmic Bayesian factor between the neutrino NH and IH scenarios is compatible with zero within one standard deviation,
while the difference of the best-fit $\chi^2$ is given by $\chi^2_{\mathrm{NH,min}}-\chi^2_{\mathrm{IH,min}}=-0.448$.
The adopted dataset is fitted nearly equally well by both NH and IH, but it can not distinguish the two scenarios.
Therefore, more experiments of higher precision in the future are needed to decisively distinguish the neutrino mass hierarchies \cite{Allison:2015qca,Abazajian:2013oma}.
In addition, comparing the Bayesian evidences of the two scenarios with that of $\nu_{\textrm{DH}}\Lambda$CDM, we find that the $\nu_{\textrm{DH}}\Lambda$CDM is positively supported by the adopted data combination. Based on the best-fit $\chi^2$, we find that the $\nu_{\textrm{DH}}\Lambda$CDM fits the data combination better, since $\chi^2_{\textrm{min}}$ in this scenario is smaller by around $2$ than those in the others.
In addition, the constraints in (\ref{eq:nh}) and (\ref{eq:ih}) are well consistent with those in Ref.~\cite{Wang:2016tsz}, which did not include the eBOSS DR14 data.
Another existing work \cite{Li:2017iur} has showed that the $95\%$ CL upper bound is $\Sigma m_\nu<0.118\textrm{eV}$ for the NH, and $\Sigma m_\nu<0.135\textrm{eV}$ for the IH. These constraints appear to be tighter than those obtained by this work. However, the neutrino mass hierarchy was parameterized in a different way from this work, and the data combination discarded the CMB lensing and the eBOSS DR14 BAO but included the redshift space distortion data.

From Fig.~\ref{fig:mnu}, we can confirm that the difference in neutrino mass constraints are mainly due to the different priors in the parameter $\Sigma m_\nu$.
Given the current data combination, the posterior PDF of $\Sigma m_\nu$ in the $\nu_{\textrm{NH}}\Lambda$CDM ($\nu_{\textrm{IH}}\Lambda$CDM) is approximately overlapped with that in the $\nu_{\textrm{DH}}\Lambda$CDM with a lower bound of $0.06\textrm{eV}$ ($0.10\textrm{eV}$) on $\Sigma m_\nu$.
In the $\nu_{\textrm{DH}}\Lambda$CDM model, the $95\%$ CL constraint on $\Sigma m_\nu$ is $0.164\textrm{eV}$ for a lower bound $0.06\textrm{eV}$, while it is $0.186\textrm{eV}$ for a lower bound $0.10\textrm{eV}$. These constraints are consistent with those in (\ref{eq:nh}) and (\ref{eq:ih}), respectively.
Therefore, the priors in $\Sigma m_\nu$ have significant influence on the constraints on $\Sigma m_\nu$, given the current data. However, $\nu_{\textrm{NH}}\Lambda$CDM and $\nu_{\textrm{IH}}\Lambda$CDM can fit the data slightly better than $\nu_{\textrm{DH}}\Lambda$CDM with non-zero priors in $\Sigma m_\nu$, since the best-fit $\chi^2$ in the former two scenarios are smaller by $3-4$ than those in the latter two.
Comparing $\nu_{\textrm{NH}}\Lambda$CDM with $\nu_{\textrm{DH}}\Lambda$CDM+0.06eV prior, we find a negative support for the former scenario due to $\Delta \ln E=-1.29$.
Comparing $\nu_{\textrm{IH}}\Lambda$CDM with $\nu_{\textrm{DH}}\Lambda$CDM+0.10eV prior, we find no significant support for either scenario due to $\Delta \ln E=-0.11$.

\section{Results for the extended cosmological models}\label{sec:result2}

\begin{figure*}
\begin{minipage}[t]{0.696\columnwidth}
\centering
\includegraphics[width=1\columnwidth,height=0.8\columnwidth]{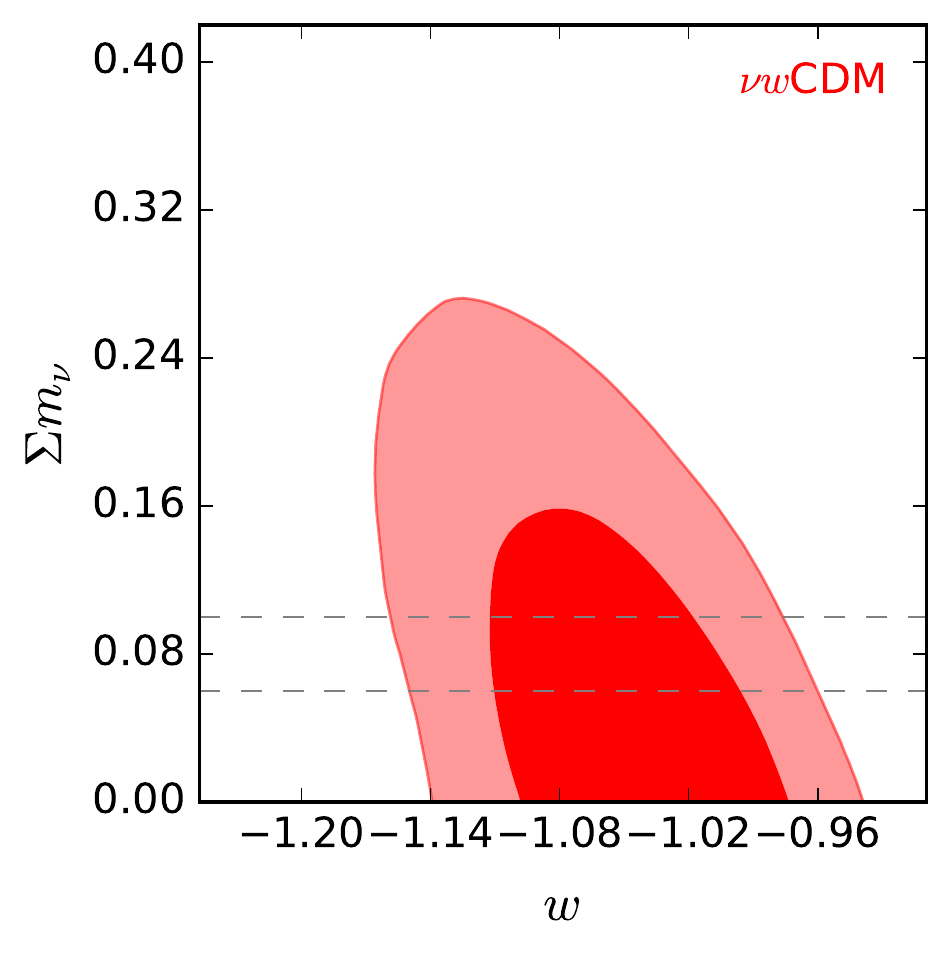}
\end{minipage}%
\begin{minipage}[t]{0.696\columnwidth}
\centering
\includegraphics[width=1\columnwidth,height=0.8\columnwidth]{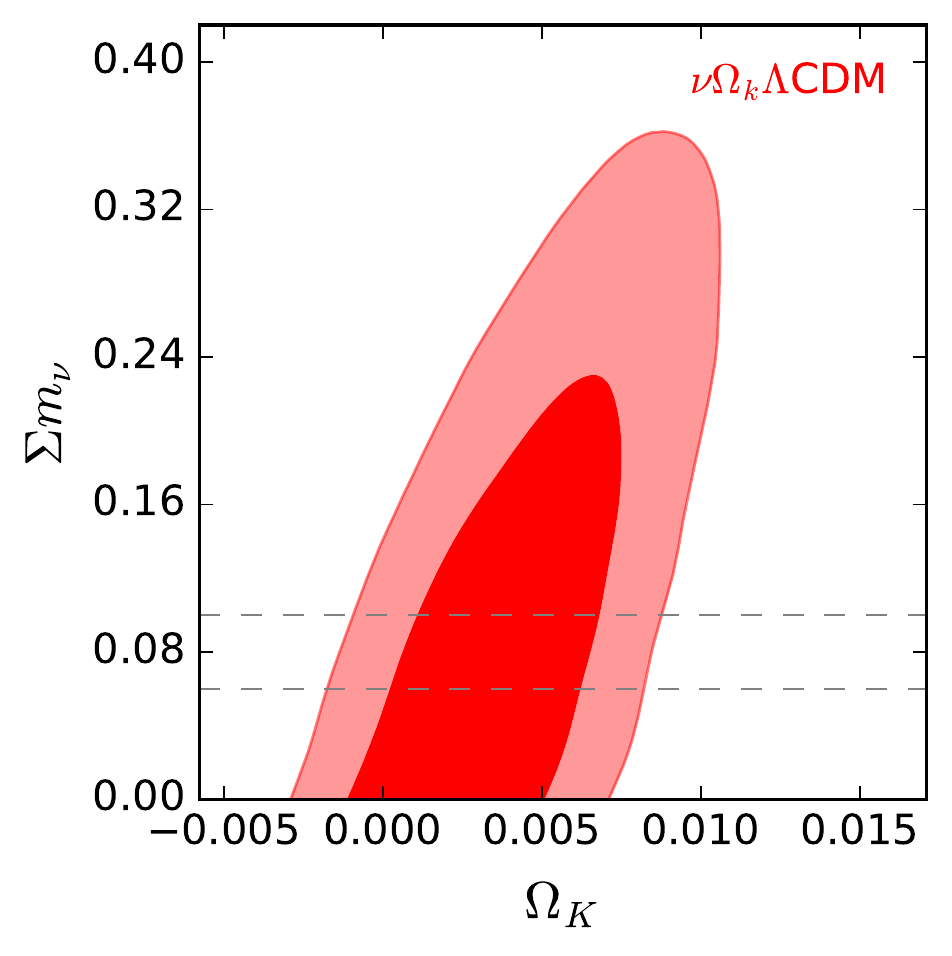}
\end{minipage}%
\begin{minipage}[t]{0.696\columnwidth}
\centering
\includegraphics[width=1\columnwidth,height=0.8\columnwidth]{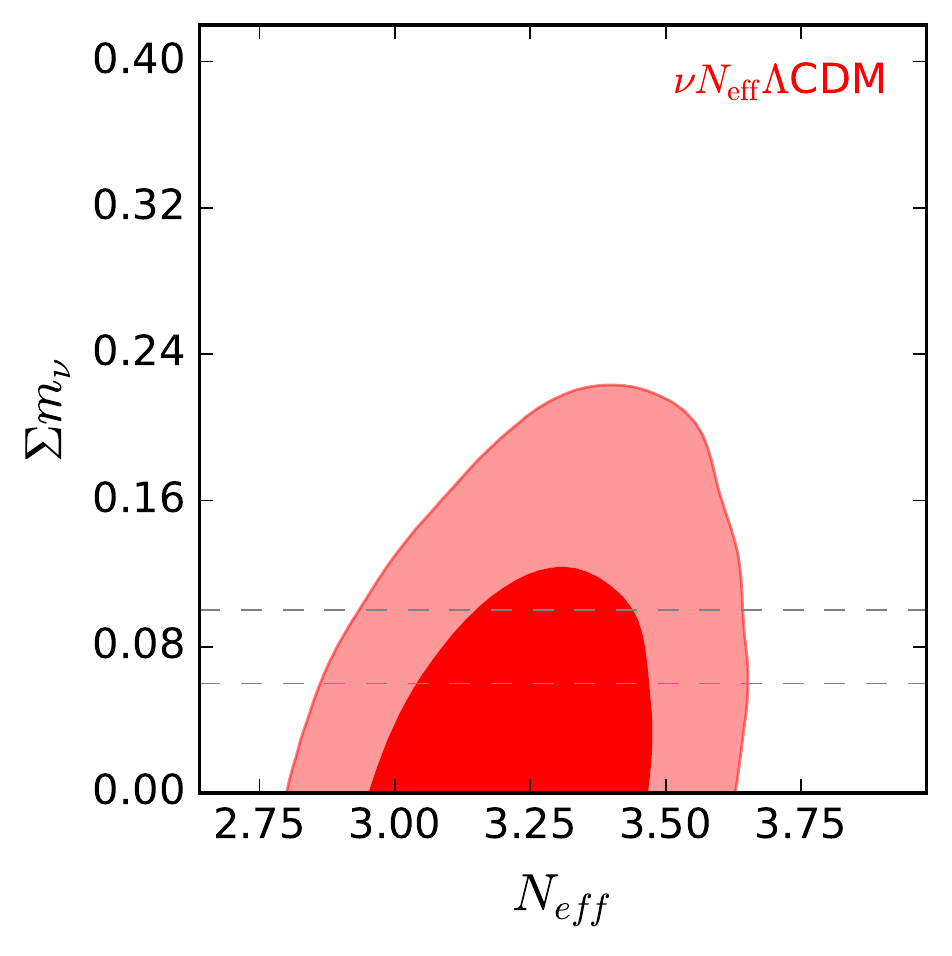}
\end{minipage}
\caption{Assuming the degenerate mass hierarchy, the $1\sigma$ and $2\sigma$ CL contours in the two-dimensional plane spanned by the sum of neutrino masses and the extended cosmological parameters in $\nu w$CDM, $\nu \Omega_k \Lambda$CDM, and $\nu N_{\mathrm{eff}}\Lambda$CDM, respectively. From bottom to up, the horizontal dashed lines denote $\Sigma m_\nu=0.06~\mathrm{eV},~0.10~\mathrm{eV}$, respectively.}
\label{fig:mnu2}
\end{figure*}

Based on the precision of current cosmological observations, the neutrino mass hierarchies have negligible effects on the constraints on $\Sigma m_\nu$ in the extended cosmological models explored here. We thus explore the parameter space by assuming the degenerate mass hierarchy in the following.
For the extended cosmological models, we present the results of our data analysis in Tab.~\ref{tab:para2} and in Fig.~\ref{fig:mnu2}. Specifically, we show the $95\%$ CL upper limits on the sum of neutrino masses, and the $68\%$ CL constraints on the remaining seven parameters in Tab.~\ref{tab:para2}. For each extended model, we depict the $1\sigma$ and $2\sigma$ \ac{CL} contours in the two-dimensional plane spanned by the sum of neutrino masses and the extended parameter in Fig.~\ref{fig:mnu2}.

\begin{table}[htbp]
\centering
\renewcommand{\arraystretch}{1.5}
\scalebox{0.89}[0.89]{%
\begin{tabular}{|c|c|c|c|c|}
\hline
& $\nu w$CDM & $\nu\Omega_k\Lambda$CDM & $\nu N_{\mathrm{eff}}\Lambda$CDM  \\
\hline
$\omega_b$ & 
 $0.02230\pm0.00015$ & $0.02222\pm0.00016$ & $0.02252\pm0.00018$ \\
$\omega_c$ & $0.1186\pm0.0012$ & $0.1198\pm0.0015$ & $0.1212\pm0.0027$ \\
$100\theta_{\textrm{MC}}$ &  $1.04093\pm0.00030$ & $1.04074\pm0.00034$ & $1.04067\pm0.00041$ \\
$\tau$ & $0.0659\pm0.0146$ & $0.0736\pm0.0159$ & $0.0734\pm0.0145$ \\
$n_s$ & $0.9670\pm0.0043$ & $0.9643\pm0.0049$ & $0.9765\pm0.0069$ \\
$\textrm{ln}(10^{10}A_s)$ & $3.063\pm0.027$ & $3.082\pm0.031$ & $3.084\pm0.029$ \\
$w$ & $-1.06_{-0.04}^{+0.05}$ & -- & -- \\
$\Omega_k$ & -- & $0.0043_{-0.0028}^{+0.0024}$ & -- \\
$N_{\mathrm{eff}}$ & -- & -- & $3.264_{-0.161}^{+0.160}$ \\
$\Sigma m_\nu~[\textrm{eV}]$ & $<0.214$ & $<0.294$ & $<0.174$ \\
\hline
$\chi^2_{\textrm{min}}/2$ & 
$6829.089$
& $6831.799$ & $6832.353$ \\
$\ln E$ & $-6892.81\pm0.24$ & $-6892.47\pm 0.24$ & $-6893.50\pm0.24$ \\
\hline
\end{tabular}}
\vspace{0.01\columnwidth}
\caption{Assuming the degenerate mass hierarchy, the $68\%$ CL constraints on the seven base parameters of the extended cosmological models, the $95\%$ CL upper limits on the sum of neutrino masses, as well as the best-fit values of $\chi^2$ and the logarithmic Bayesian evidence, i.e. $\ln E$ ($68\%$ CL).}
\label{tab:para2}
\end{table}

For the $\nu w$CDM model, we obtain the upper limit on the sum of neutrino masses to be
$\Sigma m_\nu<0.214~\mathrm{eV}$
at $95\%$ \ac{CL}. This upper limit on $\Sigma m_\nu$ is indeed improved compared with the existing ones, e.g. $\Sigma m_\nu<0.268~\mathrm{eV}$ \cite{Wang:2016tsz}, $\Sigma m_\nu<0.304~\textrm{eV}$ \cite{Zhang:2015uhk}, and $\Sigma m_\nu<0.25~\textrm{eV}$ \cite{Zhao:2016ecj}. 
The constraint on $w$ is $w=-1.06_{-0.04}^{+0.05}$ at $68\%$ \ac{CL}, deviating from $w=-1$ with a significance of 1.2$\sigma$.
From the left panel of Fig.~\ref{fig:mnu2}, $\Sigma m_\nu$ is found to be anti-correlated with $w$. Compared with the $\nu_{\mathrm{DH}}\Lambda$CDM model, we find that the logarithmic Bayesian factor is $\ln E_{\mathrm{\nu wCDM}}-\ln E_{\mathrm{DH}}=-2.31$, and the difference of the best-fit $\chi^2$ is $\chi^2_{\mathrm{\nu wCDM,min}}-\chi^2_{\mathrm{DH,min}}=-6.744$. There is thus a negative support for the $\nu w$CDM model, but this extended model fits the adopted dataset better than the $\nu_{\mathrm{DH}}\Lambda$CDM model.

For the $\nu \Omega_k \Lambda$CDM model, we obtain the upper limit on the sum of neutrino masses to be
$\Sigma m_\nu<0.294~\mathrm{eV}$
at $95\%$ CL, and the constraint on $\Omega_k$ is $\Omega_k=0.0043_{-0.0028}^{+0.0024}$ at $68\%$ CL. 
The significance of a non-zero value of $\Omega_k$ is found to be around $1.5$ standard deviations.
From the middle panel of Fig.~\ref{fig:mnu2}, $\Sigma m_\nu$ is found to be positively correlated with $\Omega_k$. Therefore, adding the parameter $\Omega_k$ worsens the constraints on the neutrino masses. 
This constraint on $\Sigma m_\nu$ is compatible with the existing one in Ref.~\cite{Chen:2016eyp}, which studied two different scenarios of neutrino mass hierarchy.
Compared with the $\nu_{\mathrm{DH}}\Lambda$CDM model, we find that the logarithmic Bayesian factor is $\ln E_{\mathrm{\nu \Omega_k\Lambda CDM}}-\ln E_{\mathrm{DH}}=-1.97$, and the difference of the best-fit $\chi^2$ is given by $\chi^2_{\mathrm{\nu \Omega_k \Lambda CDM,min}}-\chi^2_{\mathrm{DH,min}}=-1.324$. There is thus a negative support for the $\nu \Omega_k\Lambda$CDM model, even though this extended model fits the adopted dataset slightly better than the $\nu_{\mathrm{DH}}\Lambda$CDM model.

For the $\nu N_{\mathrm{eff}} \Lambda$CDM model, we obtain the upper limit on the sum of neutrino masses to be
$\Sigma m_\nu<0.17~\mathrm{eV}$
at $95\%$ CL, and the constraint on $N_{\mathrm{eff}}$ is $N_{\mathrm{eff}}=3.265_{-0.157}^{+0.159}$ at $68\%$ CL. 
Since $N_\textrm{eff}=3.046$ in standard $\Lambda$CDM model, the significance of extra relativistic degree is $1.4\sigma$ .
From the right panel of Fig.~\ref{fig:mnu2}, $\Sigma m_\nu$ is found to be positively correlated with $N_{\mathrm{eff}}$. 
This constraint on $\Sigma m_\nu$ is looser than the existing one $\Sigma m_\nu<0.14\textrm{eV}$ in Ref.~\cite{Rossi:2014nea}, which used the high-$\ell$ CMB data and the Lyman-$\alpha$ data.
Compared with the $\nu_{\mathrm{DH}}\Lambda$CDM model, we find that the logarithmic Bayesian factor is $\ln E_{\mathrm{\nu N_{\mathrm{eff}}\Lambda CDM}}-\ln E_{\mathrm{DH}}=-3.00$, and the difference of the best-fit $\chi^2$ is given by $\chi^2_{\mathrm{\nu N_{\mathrm{eff}} \Lambda CDM,min}}-\chi^2_{\mathrm{DH,min}}=-0.216$. Therefore, there is a negative support for the $\nu N_{\mathrm{eff}} \Lambda$CDM model, even though this extended model fits the adopted dataset as nearly well as the $\nu_{\mathrm{DH}}\Lambda$CDM.

\section{Conclusion and Discussion}\label{sec:summary}
In this paper, we updated the cosmological constraints on the sum of neutrino masses $\Sigma m_\nu$ using the most up-to-date observational data. 
Two more realistic mass hierarchies of massive neutrinos, namely, the normal mass hierarchy and the inverted mass hierarchy, are employed in addition to the degenerate mass hierarchy. 
In the $\nu\Lambda$CDM, for the DH, we obtained an improved upper limit $\Sigma m_\nu<0.129\textrm{eV}$ at $95\%$ \ac{CL}. 
Taking into account the squared mass differences between three massive neutrinos, we obtained the $95\%$ CL upper bound to be $\Sigma m_\nu<0.159\textrm{eV}$ for the NH, while to be $\Sigma m_\nu<0.189\textrm{eV}$ for the IH. 
Based on the Bayesian evidence, the adopted dataset can not distinguish the two mass orderings. 
In addition, we found that the priors in $\Sigma m_\nu$ can significantly impact the cosmological constraints on $\Sigma m_\nu$, given the adopted data combination.
Future cosmological observations of higher precision are needed to get more decisive conclusions. For example, BAO \cite{Zhao:2015gua,Font-Ribera:2013rwa}, CMB \cite{Calabrese:2014gwa,Benson:2014qhw,Matsumura:2013aja,Kogut:2011xw}, and galaxy shear surveys \cite{  Abell:2009aa,Laureijs:2011gra} might reach the sensitivity to measure the neutrino masses and to determine the mass hierarchy in the future.

Since the extended cosmology model can have degeneracy with the  neutrino mass, we extended our studies to include the generic dark energy, the spatial curvature, and the extra relativistic degrees of freedom, respectively, by assuming the degenerate mass hierarchy. Compared with the $\nu_{\mathrm{DH}}\Lambda$CDM model, we found negative supports for these extended cosmological models based on Bayesian model selection, due to the introduction of an additional independent parameter in each model. 

We compared the results of this work with the existing ones. Comparing with our existing works \cite{Huang:2015wrx,Wang:2016tsz}, which used the same data sets except the eBOSS DR14, we found that the eBOSS DR14 brings about at most a few percent corrections to the neutrino mass constraints. However, it is challenging to compare this work with others, since they usually used different combinations of cosmological data or even different models. For example, adding the CMB lensing to the data combination can worsen the constraints on the neutrino masses \cite{Ade:2015xua}.
When the CMB lensing was discarded for the $\nu_{\textrm{DH}}\Lambda$CDM, the $95\%$ CL upper limit on $\Sigma m_{\nu}$ even became $0.10\textrm{eV}$, as found by this work. It is much tighter than that in (\ref{eq:dh}), and has reached the minimal mass expected in the IH scenario. For a second example, adding a prior on the reionization optical depth to the data combination could tighten the constraints on the neutrino masses, see for example Refs.~\cite{DiValentino:2015sam,Allison:2015qca,Vagnozzi:2017ovm,Liu:2015txa}. In addition, the degeneracy between the hot dark matter model and the massive neutrinos has been considered in Ref.~\cite{DiValentino:2015wba}, while the impacts of dynamical dark energy model on weighing massive neutrinos have been studied in Refs.~\cite{Wang:2016tsz,Zhang:2015uhk,Zhao:2016ecj,Yang:2017amu}. We have specified comparisons between the results of this work with several existing ones in last two sections.


\vspace{0.8cm}
\acknowledgments 
\noindent
We appreciate the uses of Dr.~Ning~Wu's HPC facility, and of the HPC Cluster of SKLTP/ITP-CAS. We thank Dr.~Will~Handley for his useful suggestions on the PolyChord, and Dr.~Ke~Wang for helpful discussions. SW is supported by a grant from the Research Grant Council of the Hong Kong Special Administrative Region, China (Project No. 14301214). DMX is supported by the National Natural Science Foundation of China (Grant No. 11505018) and the Chongqing Science and Technology Plan Project (Grant No. Cstc2015jvyj40031).



\end{document}